\documentclass[twocolumn,apj]{aastex63}
\usepackage{booktabs}
\usepackage{amsmath,amsthm,amsfonts,amssymb,extarrows}
\usepackage{apjfonts}

\usepackage{xcolor}
\definecolor{mBlue}{RGB}{51, 77, 167}

\hypersetup{linkcolor=red,citecolor=blue,filecolor=blue,urlcolor=blue}

\received{{2020 August 8; revised 2021 March 27; accepted 2021 April 7; published 2021 June 4} }
\shorttitle{The Astrophysical Journal, \textbf{913}, 140 (2021)}
\shortauthors{Grilo et al.}

\begin{document}
	\title{\Large Comprehensive Laboratory Measurements Resolving the LMM Dielectronic Recombination Satellite Lines in Ne-like \ion{Fe}{17} Ions}
	

	\author[0000-0001-5777-1891]{Filipe Grilo} 
	\affil{Laboratory of Instrumentation, Biomedical Engineering and Radiation Physics (LIBPhys-UNL), Department of Physics, NOVA School of Science and Technology, NOVA University Lisbon, 2829-516 Caparica, Portugal; \href{pdamaro@fct.unl.pt}{pdamaro@fct.unl.pt}}

	\author[0000-0002-6484-3803]{Chintan Shah}%
	\affil{NASA Goddard Space Flight Center, 8800 Greenbelt Rd, Greenbelt, MD 20771, USA; \href{chintan@mpi-hd.mpg.de}{{chintan@mpi-hd.mpg.de}}}
	\affil{Max-Planck-Institut f\"ur Kernphysik, Saupfercheckweg 1, 69117 Heidelberg, Germany}
	
	\author[0000-0003-0943-2101]{Steffen K\"uhn}
	\affil{Max-Planck-Institut f\"ur Kernphysik, Saupfercheckweg 1, 69117 Heidelberg, Germany}
	\affil{Heidelberg Graduate School of Fundamental Physics, Ruprecht-Karls-Universit\"at Heidelberg, Im Neuenheimer Feld 226, 69120 Heidelberg, Germany}
	
	\author[0000-0001-7442-1260]{Ren\'e Stein\-br\"ugge}
	\affil{Deutsches Elektronen-Sychrotron DESY, Notkestra{\ss}e 85, 22607 Hamburg, Germany}
	
	\author[0000-0003-0390-9984]{Keisuke Fujii}
	\affil{Department of Mechanical Engineering and Science, Graduate School of Engineering, Kyoto University, Kyoto 615-8540, Japan}
	
	\author[0000-0002-3797-3880]{Jos\'e Marques}
	\affil{BioISI – Biosystems \& Integrative Sciences Institute, Faculdade de Ci\^{e}ncias da Universidade de Lisboa, Campo Grande, C8, 1749-016, Portugal}
	\affil{Laboratory of Instrumentation, Biomedical Engineering and Radiation Physics (LIBPhys-UNL), Department of Physics, NOVA School of Science and Technology, NOVA University Lisbon, 2829-516 Caparica, Portugal}
	
	\author[0000-0001-9136-8449]{Ming Feng Gu}
	\affil{Space Science Laboratory, University of California, Berkeley, CA 94720, USA}
	
	\author[0000-0002-5890-0971]{Jos\'e Paulo Santos} 
	\affil{Laboratory of Instrumentation, Biomedical Engineering and Radiation Physics (LIBPhys-UNL), Department of Physics, NOVA School of Science and Technology, NOVA University Lisbon, 2829-516 Caparica, Portugal}
	
	\author[0000-0002-2937-8037]{Jos\'e R. {Crespo L\'opez-Urrutia}}
	\affil{Max-Planck-Institut f\"ur Kernphysik, Saupfercheckweg 1, 69117 Heidelberg, Germany}
	
	\author[0000-0002-5257-6728]{Pedro Amaro}
	\affil{Laboratory of Instrumentation, Biomedical Engineering and Radiation Physics (LIBPhys-UNL), Department of Physics, NOVA School of Science and Technology, NOVA University Lisbon, 2829-516 Caparica, Portugal}



\begin{abstract}

We investigated experimentally and theoretically dielectronic recombination (DR) populating doubly excited configurations $3l3l'$ (LMM) in \ion{Fe}{17}, the strongest channel for soft X-ray line formation in this ubiquitous species. We used two different electron beam ion traps and two complementary measurement schemes for preparing the \ion{Fe}{17} samples and evaluating their purity, observing negligible contamination effects. This allowed us to diagnose the electron density in both EBITs. We compared our experimental resonant energies and strengths with those of previous independent work at a storage ring as well as those of configuration interaction, multiconfiguration Dirac-Fock calculations, and many-body perturbation theory. This last approach showed outstanding predictive power in the comparison with the combined independent experimental results. From these we also inferred DR rate coefficients, unveiling discrepancies from those compiled in the OPEN-ADAS and AtomDB databases. 

\end{abstract}


\section{Introduction}
\label{intro}

Iron, the heaviest among the abundant chemical elements, has strong L-shell transitions that dominate the X-ray spectra of astrophysical hot (megakelvin temperature regime) plasmas in the range of 15$-$18\,\AA. Due to its closed-shell configuration with a high ionization potential of $1260~ \mbox{ eV}$, \ion{Fe}{17} (Ne-like Fe$^{+16}$) is a very stable and abundant species under those conditions.
Collisional excitation of the $3d\rightarrow2p$ and $3s\rightarrow2p$ transitions in this ion generates the strongest observed lines in the X-ray spectra (for an overview, see \citet{Brown2008} and references therein). These, together with less intense L-shell transitions from Fe in other charge states, e.g.,~  \ion{Fe}{16} (Na-like Fe$^{+15}$), provide means for diagnosing the physical conditions of those plasmas \citep{Paerels2003}. Therefore, over many years numerous laboratory measurements have aimed at providing accurate values of the wavelengths and relative intensities of L-shell transitions in Fe\,{\textsc{XV-XIX}} \citep{may2005}, \ion{Fe}{16} \citep{Graf2009}, \ion{Fe}{17} \citep{Laming2000, Beiersdorfer2002, Beiersdorfer2004a, Brown2006, Gillaspy2011, Shah2019a}, Fe\,{\textsc {{XVIII-XXIV}}} \citep{ Brown2002,chen2006}, \ion{Fe}{21} $-$ \ion{Fe}{24} \citep{chen2005}, \ion{Fe}{24} \citep{Gu1999a,chen2002a}, and Fe\,{\textsc{xxi-xxiv}} \citep{gu2001}.
These works have revealed significant discrepancies with theory; well-known problems include the 3C/3D line ratio in \ion{Fe}{17}~\citep{Bernitt2012, Kuhn2020} and the Fe solar opacity issue~\citep{Nagayama2019}. Moreover, it is expected that updated atomic data on Fe L-shells could resolve disparities among collision models used for predicting the Fe abundance~\citep{Mao2019} in low-temperature (and low-mass) elliptical galaxies \citep{Yates2017, Mernier2018}. A recent review of astrophysical diagnostics of Fe L lines can be found in~\citet{Gu2019, Gu2020}.

Dielectronic recombination (DR) is the dominant photorecombination channel for \ion{Fe}{17} in such plasmas. In the case of DR LMM, this means the capture of an electron into a vacancy of the M shell with a simultaneous, energetically resonant electron L$-$M excitation. The resulting doubly excited state can either autoionize, resulting in resonant excitation (\citep{Shah2019a} and \citep{Takashi2017} for \ion{Fe}{17} and Fe\,{\textsc{XV-XVI}}), or radiatively decay, completing the recombination: 

\begin{eqnarray}
\text{Fe}^{16+}(1s^2 2s^2 2p^6) + e^{-} \nonumber \\
\downarrow \hspace{15mm} \nonumber \\   
\text{Fe}^{15+**}(1s^2 2l^7 3l' 3l'')  \\
\downarrow \hspace{15mm} \nonumber \\   
\text{Fe}^{15+*}(1s^2 2s^2 2p^6 3l''') + h \nu ~. \nonumber
\end{eqnarray}

Among the various processes exciting L$-$M emission, DR produces strong "satellite" transitions very close to the main lines due to the perturbation caused by the added spectator electron \citep{Dubau1980, Clementson2013}. Such lines were seen with the Chandra X-Ray Observatory in spectra from stellar coronae, like those of Capella and Procyon, and are used for plasma temperature determination \citep{Beiersdorfer2018, Gu2020}. DR also strongly influences plasma ionization equilibrium \citep{Dupree1968}. It is thus crucial to accurately know these dielectronic satellites when diagnosing temperatures using collisional-radiative models \citep{Savin2002, Dudik2019}, such as AtomDB \citep{Foster2012} and SPEX \citep{Kaastra1996}, or with the help of atomic databases like CHIANTI \citep{Dere2019} and OPEN-ADAS. \footnote{\href{https://open.adas.ac.uk}{https://open.adas.ac.uk}} 

\begin{figure}[t]
 \centering
\includegraphics[clip=true,width=1.0\columnwidth]{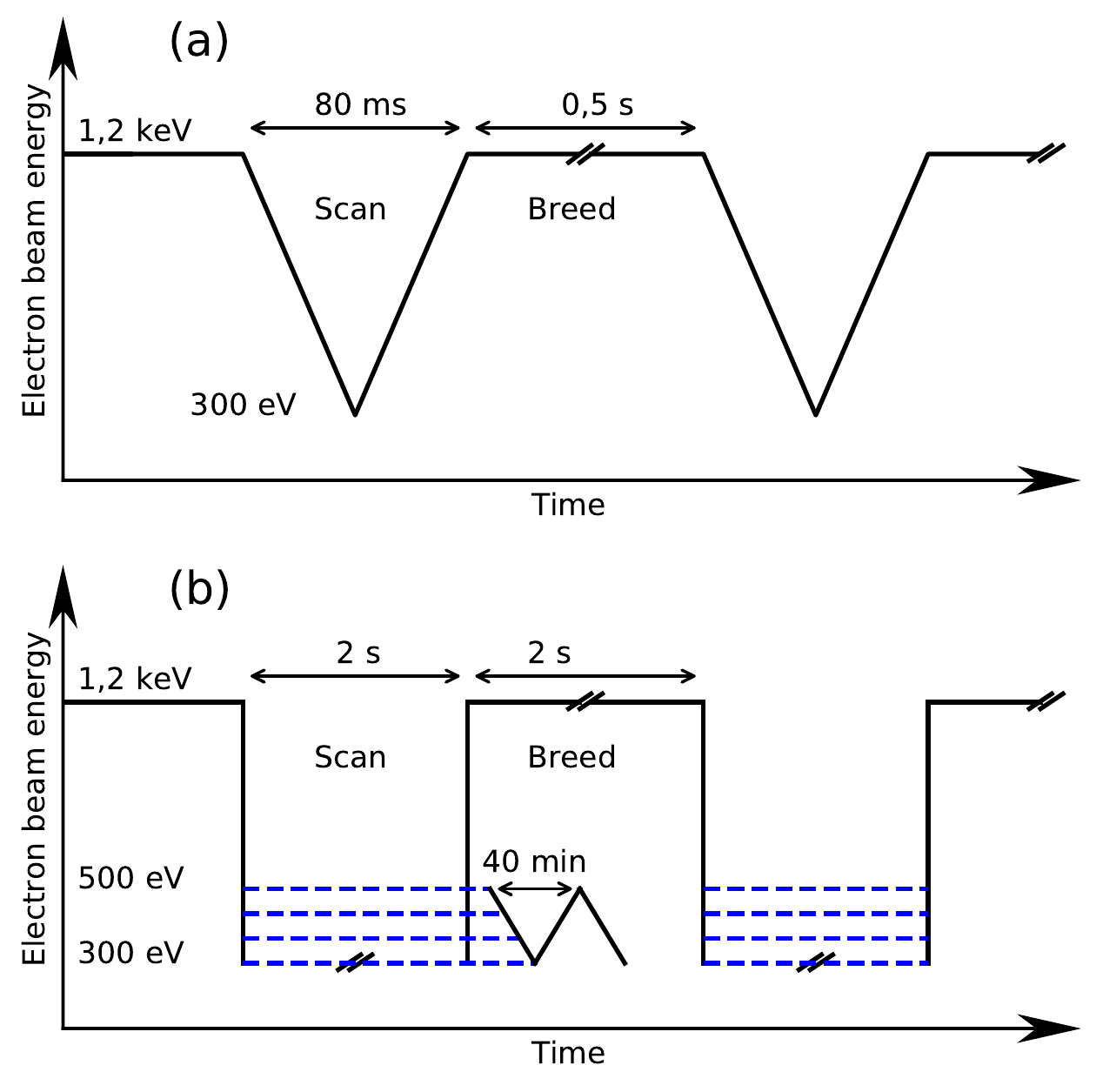}
\caption{Time pattern of the electron beam energy sweeping at (a) FLASH-EBIT \citep{Epp2010} and (b) PolarX-EBIT \citep{Micke2018}. The dashed blue lines indicate slow energy scans between 300 and 500 eV with a period of 40 minutes.} 
\label{figs:1}
\end{figure}

 Except for direct observation of DR $3l5l'$ and $3l6l'$ satellites in  Fe\,{\textsc{XXII-XXIV}} \citep{gu2001}, no laboratory wavelengths or intensities of Fe DR L-shell satellites are available, as mentioned in~\citet{Beiersdorfer2014}. Only recently have DR cross sections for the $3lnl'$ series for  \ion{Fe}{17} been published, with the purpose of investigating the $3d\rightarrow2p$ and $3s\rightarrow2p$ line ratios above the collision excitation threshold \citep{Shah2019a}. These data benchmark the SPEX model and provide constraints on the global fit of Capella spectra~\citep{Gu2020}. Continuing those works, we focus on the DR $3l3l'$ (LMM) satellites of \ion{Fe}{17}, and provide experimental resonant strengths and rate coefficients. Similar measurements were previously done for Au~\citep{Schneider1992}, Xe~\citep{DeWitt1992}, and more recently for Si~\citep{Lindroth2020}.

 In this work, we remeasured previously studied \ion{Fe}{17} LMM region by our group (see~\cite{Shah2019a}) with another electron beam ion trap (EBIT), PolarX-EBIT (described in \cite{Micke2018}). By using a modified measurement scheme, we obtained higher electron collision energy resolution compared to that in previous works. We also simulated the dynamical charge-state distribution for the present experimental conditions in order to exclude a possible large depletion of \ion{Fe}{17} ions due to DR. Furthermore, we inferred the electron beam density for both devices, obtained experimental integrated resonant strengths from the two different measurement schemes, and compared them with those of earlier photorecombination studies performed at the Heidelberg Test Storage Ring (TSR) \citep{Schmidt2009}.

In addition, our new calculations based on multiconfiguration Dirac$-$Fock (MCDF) equations and our previous ones based on the Flexible Atomic Code (FAC) were compared with configuration interaction predictions by~\citet{Nilsen1989}. 
{Finally, our experimental and theoretical resonant strengths were converted to DR rate coefficients for a few electron temperatures and compared with those available in the OPEN-ADAS database, as well as with those in~\citet{Zatsarinny2004}, which are included in AtomDB \citep{Foster2012}, a spectral modeling code widely used in the X-ray astrophysics community.}

\section{Experiment}
\label{Experi_setup}

For accurate values of DR intensities, we relied on two complementary measurements made on two different EBITs. Previous work with FLASH-EBIT at Max-Planck-Institut f\"{u}r Kernphysik in Heidelberg is reported in detail in \citet{Shah2019a}. We summarize here the method and emphasize the differences with the new measurements performed with PolarX-EBIT. In both devices, Fe atoms were injected into the trap and ionized by a magnetically compressed monoenergetic electron beam with a radius of tens of micrometers. Its negative space-charge potential confined the ions, allowing stages of high ionization to be reached. 

The DR LMM resonances appear at electron energies below the Na-like ionization threshold into Ne-like. This requires one to first produce (breed) the \ion{Fe}{17} of interest before quickly changing the interaction energy to the values to be probed, as described below. Since DR depletes the Ne-like population into Na-like if the probe time is too long, we quantify this small effect in Sec.~\ref{sec:simuresu}. Our recorded signal, X-ray emission including the contribution from DR, was observed at $90 ^\circ$ relative to the beam axis with a silicon drift detector at both EBITs. Its photon energy resolution was around $120$ eV FWHM at $6$ keV.

 \begin{figure*}[t]
 \centering
\includegraphics[clip=true,width=0.95\textwidth]{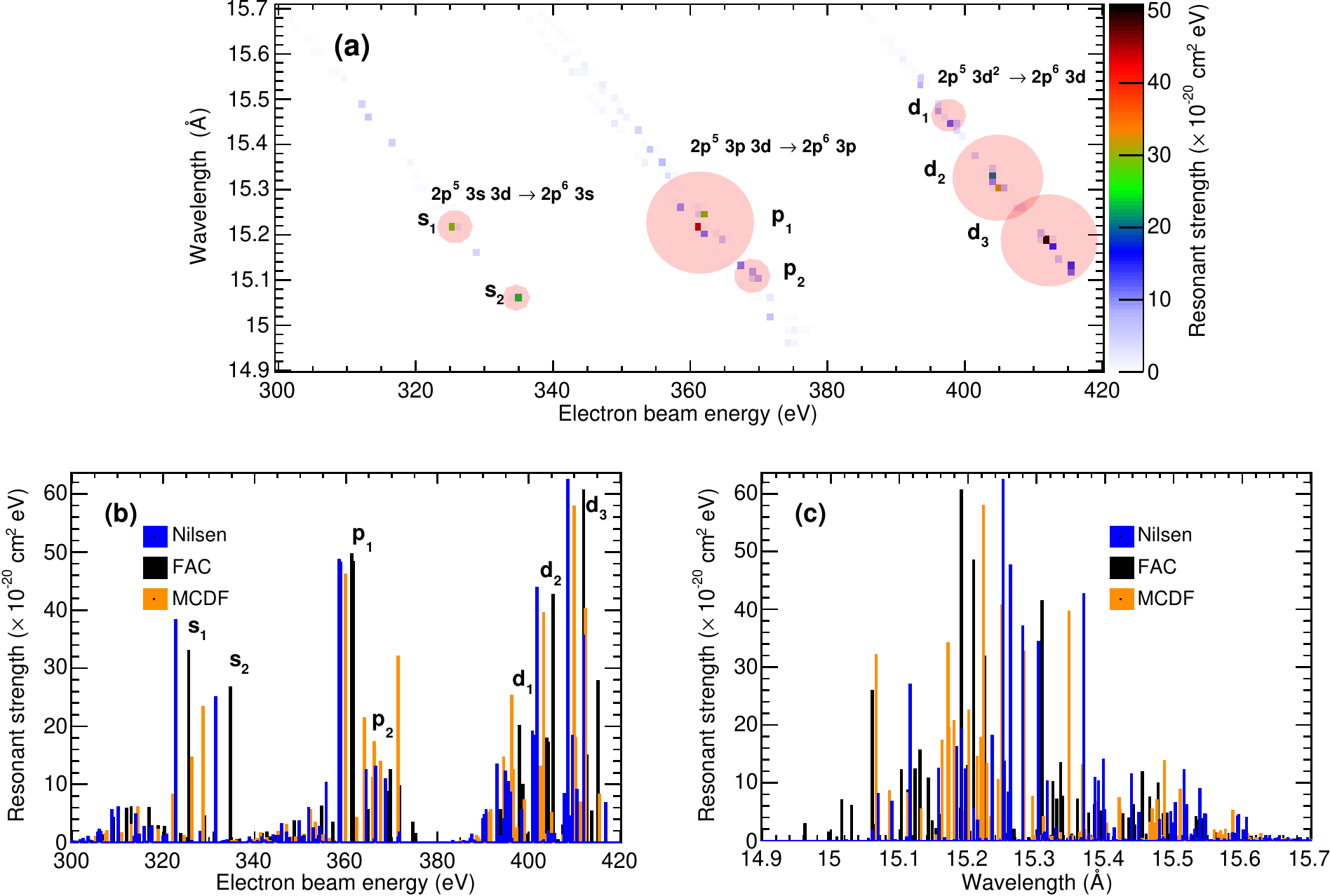} 
\caption{Top plot: Theoretical resonant strengths obtained from FAC. Abscissa: electron beam energy; ordinate: emitted wavelength. The experimentally  resolved spectral lines {(in electron beam energy)} are labelled as $s_1$, $s_2$, $p_1$, $p_2$, $d_1$, $d_2$, $d_3$ and $d_4$. The first letter of the label is the $l$ of the spectator electron in the $2p^53l3d\rightarrow 2p^63l$ transition. These lines correspond either to a single strong isolated resonance, like $s_1$, or to multiple resonances that are experimentally blended, such as $p_1$. The circle diameter shows the relative resonant strength of the channels contributing to these spectral lines. Projections onto both axes are displayed in the bottom left (b) and right (c) plots and compared with our MCDF and~\citet{Nilsen1989} calculations. The FAC and MCDF data shown in panels (b) and (c) are available as the data behind the figure in the .tar.gz package. The package also includes FAC-MBPT data. All three data sets can also be used to generate the synthetic spectra shown in Figure \ref{fig:6}. \\
(The data used to create this figure are available.)} 
\label{fig:2}
\end{figure*}

\subsection{FLASH-EBIT Measurements}
\label{sec:FLASH}

The electron beam energy was at 1.15 keV for a {0.5 s breeding time}, followed by a linear ramp-down from 1.15 to 0.3 keV within 40 ms and a symmetric ramp-up (as shown in Figure \ref{figs:1} (a); \cite{Shah2019a}), a procedure introduced by \citet{Knapp1989,Knapp1993} and used in many other experiments (e.g. \cite{Yao2010,Xiong2013,Hu2013}). For such scans, simulations of the ion population predict a negligible DR depletion of the \ion{Fe}{17} population (see Section \ref{sec:simuresu}). 

Every 5 s the ion inventory of the trap was dumped and regenerated to avoid contamination by W and Ba ions, which take typically a few minutes to accumulate. FLASH-EBIT uses superconductive coils inducing a magnetic field up to 6 T \citep{Epp2010} that strongly compresses the electron beam. This beam efficiently produces ions up to the highest charge state allowed by the ionization threshold, in this case Ne-like \ion{Fe}{17}. The residual pressure at the trap center stays below $10^{-12}$~mbar, making charge exchange (CX) with residual gas negligible. Therefore, a high-purity sample of \ion{Fe}{17} ions was prepared \citep{Shah2019a}. The beam current was adjusted according to $n_{\mbox{e}} \propto I_{\mbox{e}} / \sqrt{E}$ to keep the electronic density $n_{\mbox{e}}$ constant, having a value of 20 mA at the breeding energy. The measured electron energy spread was $\sim$5 eV.

\subsection{PolarX-EBIT Measurements}
The new measurements used PolarX-EBIT, operating at PETRA III, Deutsches Elektronen-Synchrotron (DESY), Hamburg. It uses an off-axis electron gun and a magnetic field at the trap up to 0.86 T \citep{Micke2018} produced by an assembly of permanent magnets. It operates at room temperature and had at the time of these measurements a rather poor vacuum ($10^{-8}$~mbar), which could increase the recombination from Ne-like to Na-like ions through the CX process. As observed in Section \ref{sec:simuresu}, the presence of Na-like ions is not significant. PolarX-EBIT was also run at low beam current, in the present case 2 mA, and thus reached a lower (3.5~eV FWHM at 400~eV) electron energy spread than FLASH-EBIT. The measurement scheme is different (see Figure \ref{figs:1} (b)): here, the electron beam varies according to a square-wave scan, instead of a sawtooth scan. A breeding time of 0.4 s at 1 keV is sufficient to reach \ion{Fe}{17} population equilibrium, according to both prospective measurements and simulations. The probe energy was maintained for 1 s and varied slowly (in minutes) between 300 and 500 eV. The electron energy was changed between breeding (constant) and probing values (slowly scanned) by means of a fast (tens of nanoseconds) high-voltage switch. The time evolution of the X-ray spectrum can thus be observed, in direct dependence on the evolution of the population of different charge states.

\section{Calculations}
\label{sec:theor}

DR is a resonant process involving two steps. At first, a dielectronic capture of a free electron into an initial ionic state $i$ excites a bound electron and forms a doubly excited (or intermediate) state $d$. Then, this state may radiatively decay into a final state $f$, thereby completing the DR process. 
Following our previous works \citep{Amaro2017a,Shah2018,Shah2019a}, we calculated cross sections and resonant strengths in the isolated resonance approximation, i.e., no quantum interference between DR resonances \citep{Pindzola1992}, or with nonresonant recombination channels was considered \citep{Zatsarinny2003,Martinez2005,Tu2015,Tu2016}. This contribution only influences weak resonances as been previously shown in \citet{Pindzola1992} and \citet{Zatsarinny2003}.
In this approximation, the DR strength is given by
\begin{eqnarray}
 S_{idf}^{DR} &=& \int_{0}^{\infty} \sigma^{DR}_{idf}(E_e) d E_e \nonumber \\
 			 &=& \frac{\pi^2 \hbar^3}{m_e E_{id}} \frac{g_d}{2 g_i} \frac{A^{a}_{di} A^{r}_{df}}{\sum_{i'} A^{a}_{di'} + \sum_{f'} A^{r}_{df'}},
\end{eqnarray}
where $ \sigma^{DR}_{idf}(E_e)$ is the DR cross section as a function of the free-lectron kinetic energy $E_{\mbox{e}}$. $E_{\mbox{id}}$ is the resonant energy of the electron$-$ion recombination between states $i$ and $d$, with respective statistical weights $g_{\mbox{i}}$ and $g_{\mbox{d}}$, and $m_{\mbox{e}}$ is the electron mass. The autoionization rate $A^{\mbox{a}}_{{\mbox{di}}}$ and radiative rate $A^{\mbox{r}}_{{\mbox{df}}}$ were calculated with both FAC and MCDF methods.

The details of the FAC calculation are given in \cite{Shah2019a}. FAC \citep{Gu2008} provides atomic radial wave functions and respective eigenvalues obtained in a configuration interaction method with orbitals from a modified electron$-$electron central potential. This code uses the distorted$-$wave Born approximation for calculating the autoionization rates. Beyond the standard configuration interaction module of FAC, which we employed for the initial calculations shown in Figure \ref{fig:4}, we used the many-body perturbation theory (MBPT) option of FAC \citep{gu2006} for the prediction of energies and rates in the final detailed comparisons. 

Calculations of the energies for the initial, intermediate, and final states as well as their respective transition and autoionization rates were also obtained with the Multiconfiguration Dirac$-$Fock General Matrix Elements (MCDFGME) code {(version 2008)} of Desclaux and Indelicato \citep{Desclaux1975,Indelicato1987,Indelicato1990,Desclaux2008}. Details of the method, including the Hamiltonian and the variational processes employed for retrieving wave functions, can be found in \cite{Desclaux1993} and \cite{Indelicato1995a}. In the present calculations, the electronic correlation was restricted to mixing all states of a given $j$ within an intermediate coupling (IC) scheme. 
Autoionization rates were evaluated using Fano's single-channel discrete$-$continuous expansion, which allows for non-orthogonal basis sets between the initial and final states (see~\citet{Howat1978} for details). 
Figure~\ref{fig:2} shows our calculations of DR resonant strengths obtained using FAC and the MCDF method, as well as a comparison with~\citet{Nilsen1989} values.
Note that only DR resonant strengths and the corresponding photon energies are given in \citet{Nilsen1989}. Thus, for comparing the respective resonant strengths in Figure \ref{fig:2} with our data, we inferred the corresponding resonant energies used in that work by subtracting from the photon energies the binding energy of the recombined electron in the final state. Here, the FAC-MBPT values of the binding energies were used.

Additionally, most LMM DR resonances decay through only one strong radiative channel ($2p^5 3l 3d\rightarrow 2p^6 3l$; see Figure \ref{fig:2} (a)). 
We predicted with FAC the main spectral features that can be experimentally resolved in both EBITs. We found eight spectral lines that have either a single resonant contribution, like $s_1$, or a blend of resonances, such as $p_1$ or $d_1$. Note that in our line nomenclature we use the $l$ of the $3l$ spectator electron for labeling. 
The observed line energies and strengths were compared with all theoretical predictions and available literature. Details are given in Appendix~\ref{sec:DR_cal}.

For spectral modeling, DR rate coefficients are convenient parameters. They can
be obtained by integrating the corresponding DR resonance strengths over a Maxwellian velocity distribution of the electrons \citep{Gu2003a},
\begin{equation}
 \alpha^{DR}_{if} = \frac{m_{e}}{\sqrt{\pi} \hbar^{3}} \left( \frac{4 E_{y}}{K_{B} T_e} \right)^{3/2} a^{3}_{0} \sum_{d} E_{id} S_{idf}^{DR} e^{-\frac{E_{id}}{K_{B} T_e}},
 \label{eq:DRrates}
 \end{equation}
where $E_{y}$ is the Rydberg constant in energy units, $a_0$ is the Bohr radius, $K_B$ is the Boltzmann constant, and $T_{\mbox{e}}$ is the electron temperature. 
A comparison between the present rate coefficients and those available in the OPEN-ADAS database is shown in Section \ref{sec:resso_stren}. 
%

\section{Simulations of the Charge-state Distribution}
\label{sec:simuresu}

To measure DR resonant strengths for a given ionic species, it is necessary to know the charge-state distribution of the ions trapped in the EBIT. This is mostly determined by the following charge-changing processes: collisional ionization (CI), radiative recombination (RR), DR and CX. Their competition, depending on the measurement conditions and methods, determines the overall charge-state distribution. Here, we simulate them following the work of~\citet{Penetrante1991} for computing the time evolution of the ion population in the different charge states in an EBIT by using $Z+1$ steady-state rate equations:
\begin{equation}
\begin{split}
 \frac{d N_{q}}{d t}=& n_e v_e\left[N_{q-1} \sigma_{q-1}^{C I}+N_{q+1}\left(\sigma_{q+1}^{R R}+\sigma_{q+1}^{D R}\right) \right. \\
 &\left. -N_{q} \sigma_{q}^{C I}-N_{q}\left(\sigma_{q}^{R R}+\sigma_{q}^{D R}\right) \right] \\
 &- N_{0} N_{q} \sigma_{q}^{C X} \overline{v}_{q}+N_{0} N_{q+1} \sigma_{q+1}^{C X} \overline{v}_{q+1} .
\end{split}
\label{eq:bala}
\end{equation}
Here, $N_{q}$ denotes the population of charge states $q$, $n_{\mbox{e}}$ the electron density, $v_{\mbox{e}}$ the free-electron velocity, $\sigma$ the cross section associated with a specific atomic process, and $\Bar{v}_{q}$ the mean (Maxwellian) velocity of an ion with charge $q$. 
The RR total cross sections for Mg-like, Na-like, and Ne-like Fe ions under the present experimental conditions were obtained using FAC, taking into account the principal quantum numbers up to $n~=~15$. For other Fe charge states, we used the analytical equation of~\citet{Kim1983} to obtain RR cross sections. 
CI cross sections from the measurements performed at the TSR \citep{Linkemann1995a, Hahn2013, Bernhardt2014} for Mg-like, Na-like, and Ne-like Fe ions were also used. These included, apart from the usual direct CI channel, resonant ionization processes, such as excitation and subsequent autoionization, which became strong starting from the collision excitation threshold ($\sim$750~eV). 
For the remaining charge states, CI cross sections were estimated using the Lotz formula \citep{Lotz1968a}. 
As for the CX cross sections used in our simulations, we applied the analytical expression from~\citet{Janev1983} to obtain them. 
{However, for the more abundant charge states of Ne-like, Na-like, and Mg-like Fe, we used CX cross sections obtained using the multichannel Landau$-$Zener method as implemented in FAC. An ion temperature of $\sim$60 eV for the CX cross sections was assumed, based on the previous measurement by \citet{Schnorr2013} at FLASH-EBIT for Al-like Fe. 
}  
The CX rate is proportional to the residual gas density within the trap region. Thus, by reducing the flow of the iron pentacarbonyl molecular beam in our experiment, we could enhance the population of Fe in higher charge states.
\begin{figure*}[t]
\centering
\includegraphics[clip=true,width=1.0\textwidth]{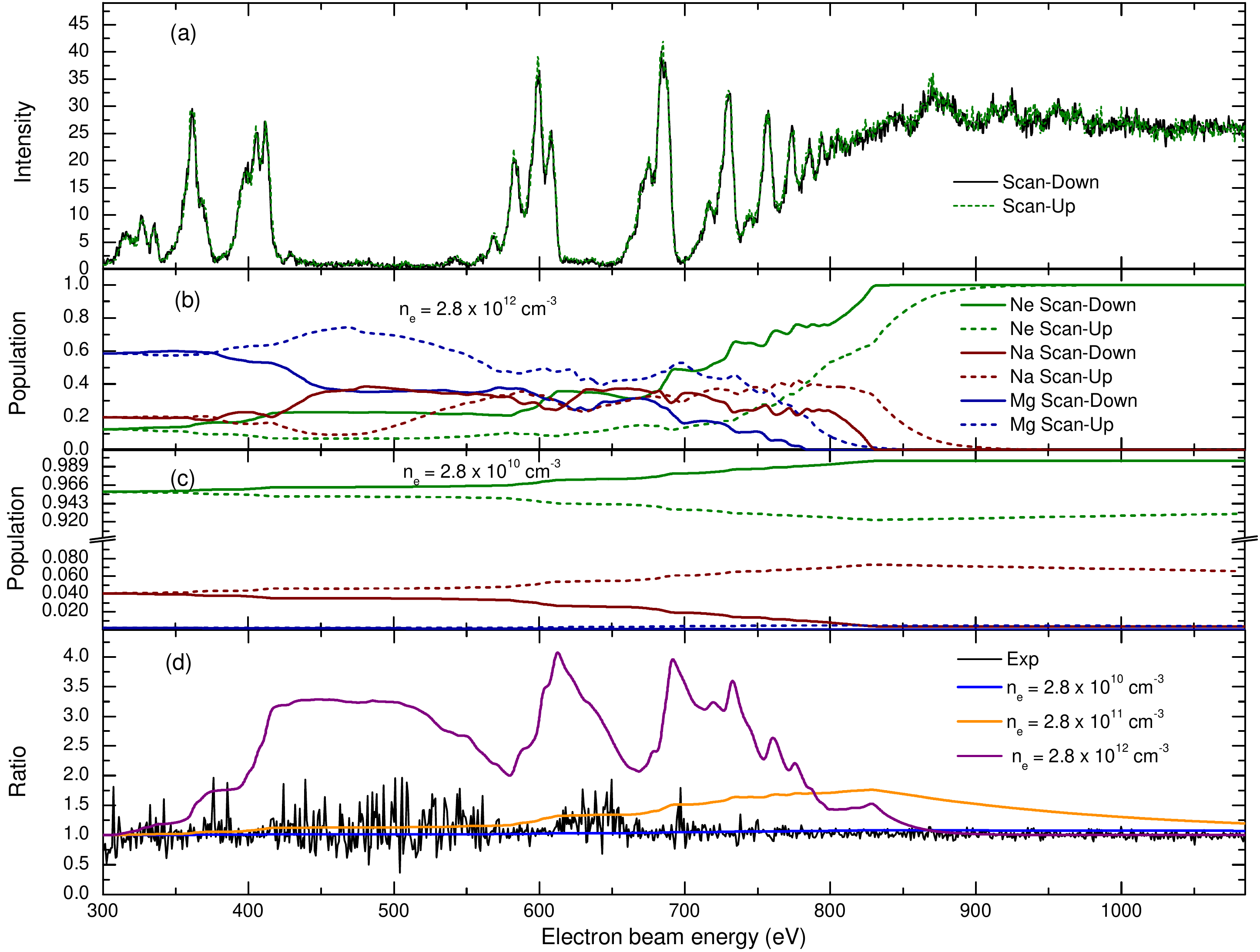}
\caption{ 
(a) Spectra observed at FLASH-EBIT by scanning upward and downward the electron beam energy (see Figure \ref{figs:1}(a) for the timing pattern). 
Simulations of Fe charge-state distributions as a function of the electron beam energy for effective electron densities of (b) { $n_{\mbox{e}}$ =
$2.8 \times 10^{12}~\mbox{cm}^{-3}$ ($I_e = 20~\mbox{mA}$) and (c) $n_e = 2.8 \times 10^{10}~\mbox{cm}^{-3}$ ($I_e = 0.2~\mbox{mA}$). The simulations were performed for an electron beam radius of $30~\mu \mbox{m}$}.  
(d) Experimental X-ray intensity ratio between downward and upward scans compared with simulated X-ray intensity ratios for different effective electron densities.
} 
\label{fig:3}
\end{figure*}

For these experiments, it is very important to choose the ratio of ionization time to recombination time appropriately to the simulation parameters. 
DR is a very strong resonant process; within the resonant width it has cross sections orders of magnitude higher than those of other collisional processes. This means that for our measurement scheme the Ne-like population in the trap should not be significantly depleted toward lower charge states by the required electron beam energy across the DR resonances. Therefore, we performed simulations for quantifying this depletion. We calculated the corresponding DR rates for Mg-like, Na-like, and Ne-like Fe ions using FAC. The principal quantum numbers of the recombined state up to $n~=~30$ for Ne-like Fe and up to $n~=~10$ for all other relevant ions were taken into account in our calculations, as well as radiative cascades for all relevant atomic processes. With these calculations we then generated synthetic X-ray emission spectra.

Furthermore, our charge balance simulations were restricted to electron densities below $10^{13}$ cm$^{-3}$, the relevant range for EBIT plasmas. At values higher than those of our simulations, we note that the effect of DR suppression has to be taken into account (see \citet{Nikolic2013, Nikolic2018} and references therein).

\subsection{FLASH-EBIT: Simulated Charge-state Distributions}
Using Equation \ref{eq:bala} and the scanning parameters presented in Figure \ref{figs:1}, we simulated the time evolution of the charge-state distribution during the upward and downward electron beam energy scans in our FLASH-EBIT measurements \citep{Shah2019a} (see Figure \ref{fig:3}(a)).
First, we investigated the effect of the electron beam density on the Ne-like, Na-like, and Mg-like Fe trapped-ion populations. 
Panel (b) of Figure \ref{fig:3} shows an extreme case of electron densities of $n_e\sim10^{12}$~cm$^{-3}$, where the Ne-like population is drastically depleted into the Na-like and Mg-like ones due to strong DR resonances. In contrast, at $n_e\sim10^{10}$~cm$^{-3}$, this effect is found to be negligible (panel (c)). 
In panel (d) of Figure \ref{fig:3}, we display the experimental ratio between the fast upward and downward energy scans, which shows a negligible effect of the scanning direction on \ion{Fe}{17} ion density in the trap. 
This stands in contrast to the out-of-equilibrium slow DR scans, where the ion populations have enough time for the decay during the scan. In this case, the two scanning directions can show clear differences in the distribution of charge states \citep{Shah2016}. 
We compared our present experimental ratio with those of simulations for different electron densities and found only a negligible charge-state depletion due to LMM DR resonances at electron densities below $5\times10^{10}$ $\text{cm}^{-3}$. 

\begin{figure}[t]
\centering
\includegraphics[clip=true,width=1.0\columnwidth]{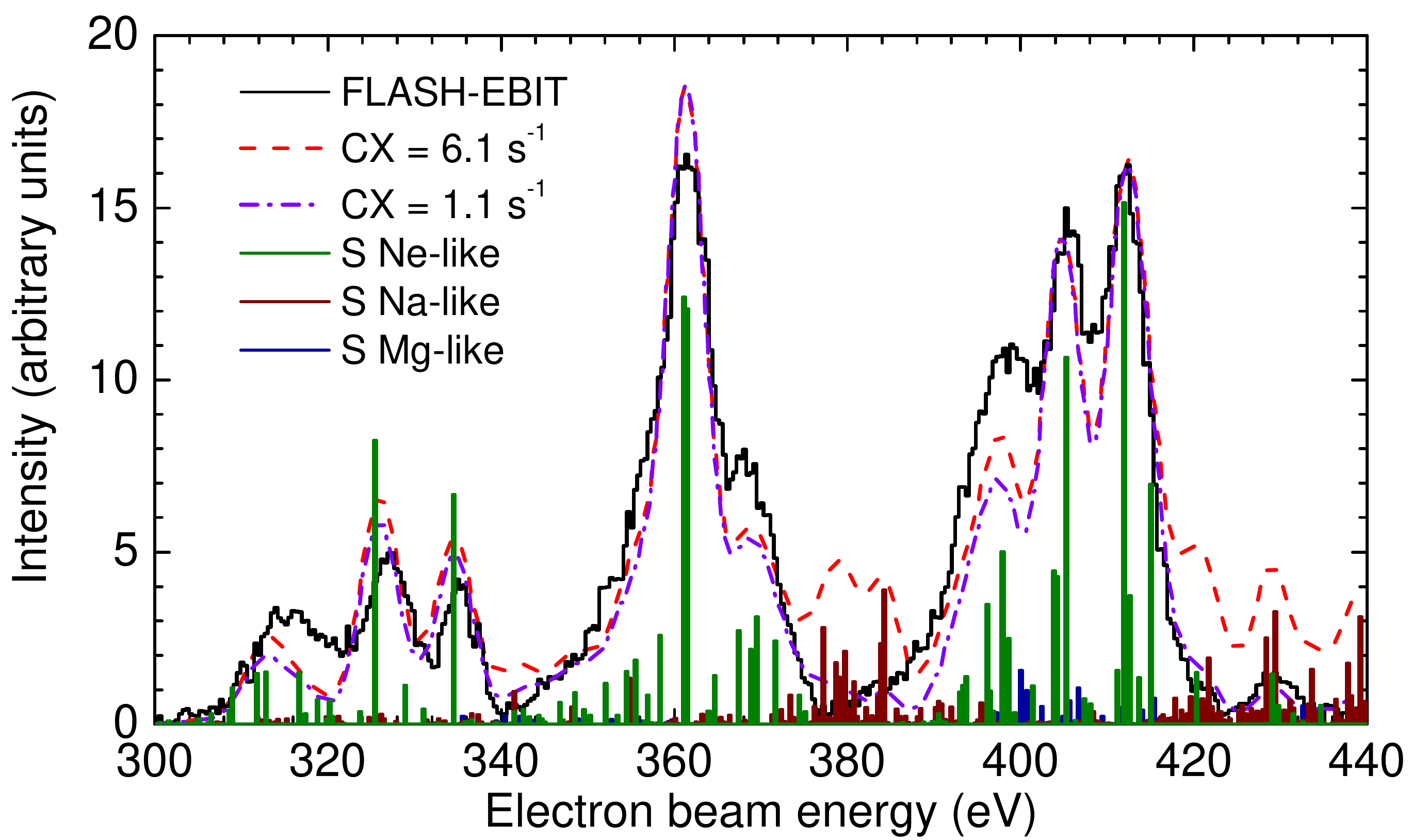}
\caption{Theoretical resonant strengths (FAC) for Ne-like, Na-like, and Mg-like Fe ions and simulated fluorescence yield under CX rates of $6.1~\mbox{s}^{-1}$ and $1.1~\mbox{s}^{-1}$ in comparison with FLASH-EBIT data. 
 } 
\label{fig:4}
\end{figure}

Second, we investigated the influence of CX on the \ion{Fe}{17} ion population distribution. FLASH-EBIT has a four-stage differential pumping system for injecting an atomic or molecular beam into the trap, where the ions are generated.
The first two stages operate at room temperature at pressures of $\sim 8 \times 10^{-9}$ mbar. Two additional stages operate cryogenically at 45\,K and 4\,K, and further constrain the gas flow into the trap region. 
This brings the residual gas pressure well below $\leq 10^{-11}$ mbar at the trap center and tremendously reduces the CX rates.
For the study of a possible influence of CX in our measurements, we simulated the ion population and generated synthetic X-ray spectra for CX rates of 0.23 s$^{-1}$ and 0.05 s$^{-1}$ (see Figure \ref{fig:4}).
When Fe XVI and Fe XV ions are produced by CX, distinct DR resonances of these ions appear at beam energies of 380 and 440 eV. Since those resonances were not observed, we conclude that under the present conditions the dominant Ne-like Fe ion population was maintained during the FLASH-EBIT measurements \citep{Shah2019a}.

\subsection{PolarX-EBIT: Simulated Charge-state Distributions}
\label{sec:simuPolar}


\begin{figure}
\centering
\includegraphics[clip=true,width=1.0\columnwidth]{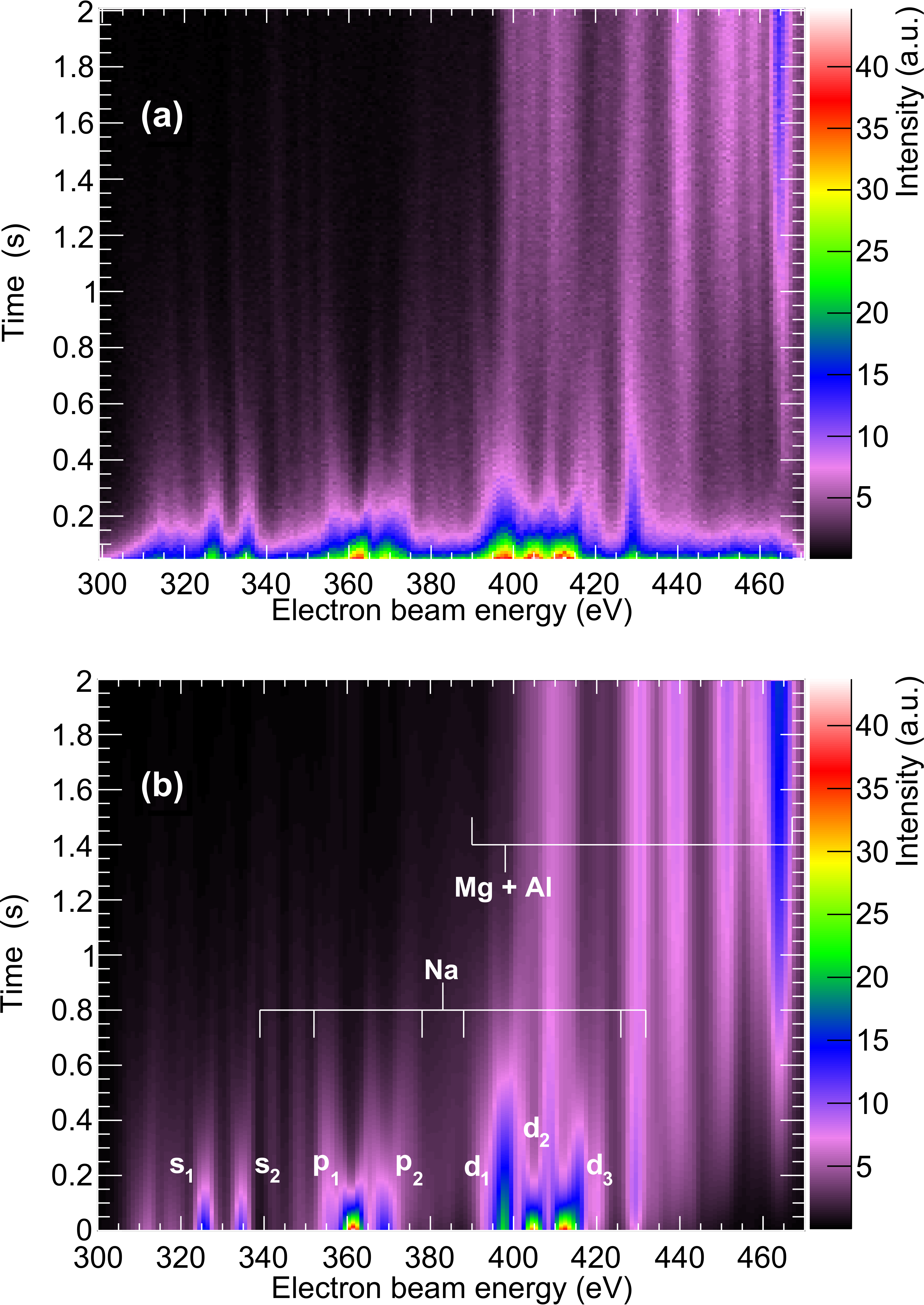}
\caption{(a) Measured fluorescence yield as a function of electron beam energy and probing time at PolarX-EBIT. (b) Simulation of the DR emission at the experimental conditions.} 
\label{fig:5}
\end{figure}

We also simulated the time evolution of the charge-state distribution and its effect on the observed LMM DR X-ray emission for PolarX-EBIT conditions, according to the measurement technique shown in Figure \ref{figs:1}(b). 
Here we compare the measurement and simulation in Figure \ref{fig:5}. 
{
For probing times shorter than 100 ms, we observed a dominant population of Ne-like ions with a small population of Na-like and Mg-like Fe ions. 
However, the observed intensities of Na-like and Mg-like LMM resonances are slightly higher than those in the FLASH-EBIT measurements (see Figure \ref{fig:5} and Figure \ref{fig:4} for the resonant energy positions). 
Simulations at the breeding electron beam energy of 1.2 keV used in the PolarX-EBIT measurements yielded populations of {0.92$\pm$0.04}, {0.07$\pm$0.03}, and {0.003$\pm$0.002} for Ne-like, Na-like, and Mg-like Fe ions, respectively.
These values are consistent with a negligible CX rate of $0.47~\mbox{s}^{-1}$, as likewise observed in the FLASH-EBIT measurements. 
}

{
The uncertainties given for the ion populations were estimated based on the difference between the cross sections obtained with the analytical formula of a given process (e.g., from ~\cite{Kim1983} for the RR case) and those of FAC calculations. 
As for the CI rates, besides using the difference between FAC and~\citet{Lotz1968a} values, we also took into account the experimental uncertainties from the CI measurements performed at the TSR \citep{Linkemann1995a, Hahn2013, Bernhardt2014} for Mg-like, Na-like, and Ne-like Fe ions.
For the CX cross sections, we found relative deviations between the expression of~\citet{Janev1983} and FAC as high as 40\%, which is in accord with observations of \citet{Betancourt-Martinez2018}. 
However, due to the very low CX rate, uncertainties associated with the CX cross sections and ion temperature do not significantly affect the predicted Ne-like populations.
}

The most intense Ne-like resonances exhibit decay times between 0.07 and 0.13 s in our experiment, while Na-like resonances (e.g., at 345, 380, and 435 eV beam energies) reach their maxima between 0.2 and 0.4 s as the Ne-like population starts to deplete.
For probing times longer than 1\,s, the spectra are dominated by DR emission from the Mg-like and Al-like Fe populations. 
We also observe a constant X-Ray emission background in our experiment, which cannot be explained by simulations under any conditions of electron density and CX rate. 
It might be attributed to the delayed photon emission from metastable states. However, no exponential decay of the signal was observed in the FLASH-EBIT data. Another possible source might be the electronic analog-to-digital converter (ADC) noise caused by switching between power supplies. 

Finally, when considering various recombination data either from FAC or MCDF theories or from TSR measurements in the charge-state distribution simulations, we did not see any change in the final synthetic spectrum. 
Also, this does not change our conclusion that neither spurious charge states other than Ne-like Fe nor charge-state depletion due to DR significantly affects any of the two measurements.

\section{Data Analysis and Results}
\label{sec:data_analy_results}

\begin{figure}
\centering
\includegraphics[clip=true,width=\columnwidth]{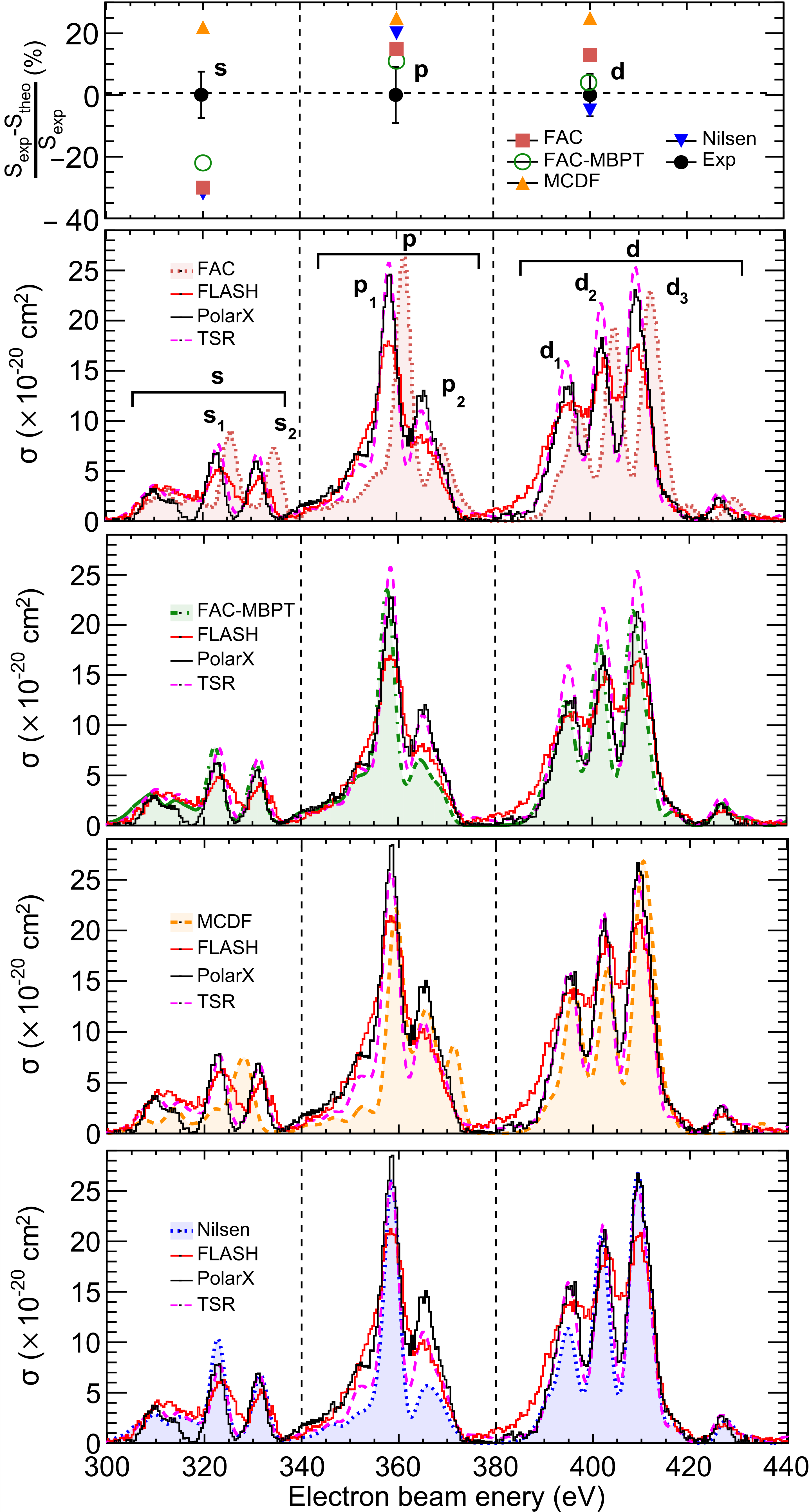} 
\caption{
\ion{Fe}{17} LMM DR resonance strengths measured using FLASH-EBIT, PolarX-EBIT and TSR, compared with the theoretical total DR cross sections obtained using FAC, FAC-MBPT, MCDF, and literature values from~\citet{Nilsen1989}. The top panel compares the averaged integrated resonance strengths from three measurements for the $s$, $p$, and $d$ regions with different theories. Note that the calculations were convoluted with a Gaussian of {3.5} eV FWHM for the comparison and FLASH-EBIT and PolarX-EBIT measurements were calibrated to each of the theoretical cross sections. 
} 
\label{fig:6}
\end{figure}

Experimental spectra observed in the FLASH-EBIT, PolarX-EBIT, and TSR measurements are shown in Figure \ref{fig:6}.
In the case of PolarX-EBIT, as explained in Section \ref{sec:simuresu}, we have only selected the first 50 ms of data in order to avoid possible charge-state depletion due to the DR and CX processes (see Figure \ref{fig:5}).
Strong  DR lines were identified in Section \ref{sec:theor}, which are clearly resolved with an excellent collision energy resolution provided by both EBITs. 
The TSR data have a significantly better resolution than the EBIT results. 
However, for our comparison we synthetically broadened the TSR spectrum to match the PolarX-EBIT spectral resolution (for a complete data set, see \citet{Schmidt2009,Shah2019a}). 

We calibrated the electron beam energy axis using the TSR data of the $p_1$ and $d_3$ peaks. 
In order to compare the resonance strengths, we first subtracted an assumed linear background from the RR continua (RR+ADC for PolarX, see Section \ref{sec:simuPolar}). 
The observed photon intensities in both EBIT spectra were normalized to a DR resonance $d_3$ at 412 eV of electron beam energy. 
Details of such cross-section normalization are explained in our previous work \citep{Shah2019a}. 
In EBITs, the unidirectional electron beam causes anisotropic and polarized X-ray emission. We observed photons at $90 ^\circ$ with respect to the electron beam. 
We thus applied a correction factor defined as $W(90 ^\circ) = 3/(3 - P)$, where $P$ is the polarization for a specific radiative transition \citep{Shah2018}. The calculated polarization values were taken from~\citet{Shah2019a}, which also agree with measurements performed by~\citet{chen2004pol}.
With this correction, we obtained the total cross sections using the $S^\mathrm{total} = 4\pi\,\, I^{90^\circ}/ W(90^\circ)$ relation, where $I^{90^\circ}$ is the observed DR intensity. 
As observed in Figure \ref{fig:4} (Ne-like case), there are more than $\sim$200 LMM DR resonances within the scanned electron beam energy range. 
As most of the features observed in Figure \ref{fig:6} consist of a number of blended resonances, we only provide the integrated resonance strengths. For this, we defined three energy regions: $s$, 300 eV to 335 eV; $p$, 335 eV to 380 eV; and $d$, 380 eV to 440 eV. 
%

\subsection{DR Energies and Strengths}
\label{sec:resso_stren}

Figure~\ref{fig:6} shows a comparison between the experimental data and predictions by FAC, FAC-MBPT, MCDF, and \citet{Nilsen1989}. 
All theories except MCDF predict all experimental features. 
FAC, in particular, shows a systematic shift in resonance energies as compared with the observed data and other theories. 
Also, disagreements of MCDF predictions are clearly visible for the $s_1$, $s_2$, and $p_2$ features. 
DR resonances within these features show larger energy splitting in the MCDF predictions than in the FAC and FAC-MBPT ones. 
Moreover, a clear feature at 425 eV is not predicted by MCDF. 
The reason behind this might be that the MCDFGME package \citep{Desclaux2008} used in the present work could not generate reliable Auger rates with correlation to other configurations beyond minimal coupling, which would be necessary to improve the energy centroid accuracy. 

Table~\ref{table:2} presents experimentally inferred {integrated} resonant strengths for the $s$, $p$, and $d$ regions. 
We estimated total uncertainties on the level of $\sim$13\% and $\sim$11\%, respectively, for FLASH-EBIT and PolarX-EBIT integrated resonance strengths, mainly arising from counting statistics and cross-section normalization. 
For the TSR results, we used the quoted uncertainty of 20\% from the original work of~\citet{Schmidt2009}. 
By combining these three independent measurements, we obtained final values for the integrated resonant strengths $S_\mathrm{EXP}$ and their respective uncertainties, found to be at the level of $\sim$10\% (see Table \ref{table:2}).
As shown in the top panel of Figure \ref{fig:6}, FAC-MBPT shows good agreement with the inferred integrated strengths, except for region $s$. 
Interestingly, all theories disagree with our inferred resonance strength for the $s$ region.
We also note that no calculations can effectively predict the resonant strength of the $p_2$ feature. 
EBIT and TSR data also disagree here; since simulations seem to exclude spurious features at this energy in EBITs, at present we do not have an explanation for this.

%
\setlength{\tabcolsep}{6pt}
\renewcommand{\arraystretch}{1.4}
\begin{table*}
\centering
\caption{Experimental Integrated Resonant Strengths ($10^{-20} \mbox{cm}^{2} \mbox{eV}$) Compared to Values Obtained with FAC and MCDF (This Work) and Those Reported by~\citet{Nilsen1989}. }
\begin{tabular*}{\textwidth}{c @{\extracolsep{\fill}} clccccccc}
\hline
\hline
Label	&	$S_{\text{FLASH}}$	&	$S_{\text{PolarX}}$	&	$S_{\text{TSR}}$	&	$S_{\text{EXP}}$	&	$S_{\text{FAC}}$	&	$S_{\text{FAC-MBPT}}$	&	$S_{\text{MCDF}}$	& \citet{Nilsen1989}	\\
\hline
{$s$}       &   {$80 \pm 10$}    &  { $65 \pm 9$ }   &   {$100 \pm 10$}    &     { $80 \pm 6$}      &     {104.27 (-30\%) }   &    {97.77 (-22\%) }  &    { 62.66  (22\%)}   &      {106.25 (-32\%)}     \\
{$p$}       &   {$210 \pm 30$}    &  { $230 \pm 20$ }   &   {$230 \pm 50$}    &     { $220 \pm 20$}      &     {201.18 (15\%) }   &    {198.34 (11\%) }  &    {202.89   (25\%)}   &      {174.30 (20\%)}     \\
{$d$ }      &     {$300 \pm 40$ }  &   {$280 \pm 30$}    &    {$330 \pm 70$}     &     {$290 \pm 20$}     &     { 271.76   (13\%)}   &      {283.43 (4\%)}      &      {270.29    (25\%)}   &     {306.67   (-5\%)}   \\
\hline
\end{tabular*}\\
\vspace{1ex}
{\raggedright \textbf{Note.} The listed experimental cross sections were calibrated with the FAC-MBPT theoretical value of the d$_3$ feature. Round brackets: Relative difference in percent between measured and theoretical resonant strengths, where each measurement was independently calibrated with the $d_3$ value from respective theories. The labels $s$, $p$ and $d$ correspond to  regions of $300~\mbox{eV}$ to $340~\mbox{eV}$, $340~\mbox{eV}$ to $380~\mbox{eV}$, and $380~\mbox{eV}$ to $420~\mbox{eV}$, respectively. \par}

\label{table:2}
\end{table*}


\setlength{\tabcolsep}{4pt}
\renewcommand{\arraystretch}{1.0}
\begin{table*}
\centering
\caption{{Experimental DR Rate Coefficients ($ \times 10^{-13} \mbox{cm}^{3} \mbox{s}^{-1}$) for a Few Electron Temperatures $T_{\mbox{e}}$ (eV) Compared to Values Obtained with FAC-MBPT and MCDF and by~\citet{Nilsen1989} and by~\citet{Zatsarinny2004}.}}
{
\begin{tabular*}{\textwidth}{c @{\extracolsep{\fill}} ccccccccccc}
\hline
\hline
Label	&	$T_{\mbox{e}}$	&	Exp.	&	\multicolumn{4}{c}{OPEN-ADAS} &	
\multicolumn{1}{c}{\hspace{0.3cm}FAC-MBPT~}	&	\multicolumn{1}{c}{MCDF~}	&	\multicolumn{1}{c}{ \citet{Nilsen1989}~}	&	\multicolumn{1}{c}{~\citet{Zatsarinny2004}}\footnote{ Only total DR LMM rates are provided.}	\\
\cline{4-7}
    &   &   &   oizLS  &   oizIC  &   nrbLS  &   nrbIC  &  & & &  \\
\hline
$3s$	&	110.3	&	$\cdots$	&	10.9	&	9.24	&	10.6	&	7.04	&	8.96	&	6.40	&	9.28	&	$\cdots$	\\
	&	220.3	&	$\cdots$	&	17.4	&	15.0	&	17.4	&	11.9	&	13.7	&	10.3	&	14.4	&	$\cdots$	\\
	&	300	&	$\cdots$	&	$\cdots$	&	$\cdots$	&	$\cdots$	&	$\cdots$	&	12.8	&	9.71	&	13.5	&	$\cdots$	\\
	&	2000	&	$\cdots$	&	$\cdots$	&	$\cdots$	&	$\cdots$	&	$\cdots$	&	1.89	&	1.48	&	2.00	&	$\cdots$	\\
$3p$	&	110.3	&	$\cdots$	&	20.1	&	20.4	&	16.8	&	17.2	&	18.8	&	17.1	&	16.5	&	$\cdots$	\\
	&	220.3	&	$\cdots$	&	40.3	&	42.2	&	36.4	&	38.9	&	33.0	&	30.9	&	29.5	&	$\cdots$	\\
	&	300	&	$\cdots$	&	$\cdots$	&	$\cdots$	&	$\cdots$	&	$\cdots$	&	31.9	&	30.0	&	28.6	&	$\cdots$	\\
	&	2000	&	$\cdots$	&	$\cdots$	&	$\cdots$	&	$\cdots$	&	$\cdots$	&	5.12	&	4.86	&	4.63	&	$\cdots$	\\
$3d$	&	110.3	&	$\cdots$	&	18.9	&	21.2	&	15.9	&	16.9	&	19.2	&	16.8	&	20.6	&	$\cdots$	\\
	&	220.3	&	$\cdots$	&	46.2	&	52.6	&	41.4	&	45.4	&	40.7	&	36.6	&	43.8	&	$\cdots$	\\
	&	300	&	$\cdots$	&	$\cdots$	&	$\cdots$	&	$\cdots$	&	$\cdots$	&	41.3	&	37.5	&	44.5	&	$\cdots$	\\
	&	2000	&	$\cdots$	&	$\cdots$	&	$\cdots$	&	$\cdots$	&	$\cdots$	&	7.46	&	6.86	&	8.06	&	$\cdots$	\\
\hline
total	&	110.3	&	$47 \pm 3$	&	49.9	&	50.9	&	43.3	&	41.2	&	46.9	&	40.3	&	46.4	&	46.6	\\
	&		&		&	 (-6.2\%)	&	 (-8.2\%)	&	 (7.9\%)	&	 (12\%)	&	 (0.2\%)	&	 (14\%)	&	 (1.3\%)	&	 (0.8\%)	\\
	&	220.3	&	$88 \pm 4$	&	103.9	&	109.8	&	95.2	&	96.2	&	87.4	&	77.7	&	87.6	&	91.0	\\
	&		&		&	 (-18\%)	&	 (-25\%)	&	 (-8.2\%)	&	 (-9.3\%)	&	 (0.7\%)	&	 (12\%)	&	 (0.4\%)	&	 (-3.5\%)	\\
	&	300	&	$87 \pm 4$	&	$\cdots$	&	$\cdots$	&	$\cdots$	&	$\cdots$	&	86.0	&	77.2	&	86.6	&	90.3	\\
	&		&		&	& & &	&	 (0.8\%)	&	 (11\%)	&	 (0.1\%)	&	 (-4.2\%)	\\
	&	2000	&	$14.5 \pm 0.7$	&	$\cdots$	&	$\cdots$	&	$\cdots$	&	$\cdots$	&	14.5	&	13.2	&	14.7	&	15.3	\\
	&		&		&	& & &	&	 (0.3\%)	&	 (9.3\%)	&	(-1.0\%)	&	 (-5.2\%)	\\
\hline
\end{tabular*}\\
\vspace{1ex}
{\raggedright \textbf{Notes.} The first label indicates the orbital of the decoupled electron in the final state after DR emission. { Labels oizLS, oizIC, nrbLS and nrbIC refer to data files available at \texttt{adf09} of OPEN-ADAS}. \par}
}
\label{table:3}
\end{table*}

\subsection{DR Rate Coefficients}
\label{sec:DR_coeff}
{
In addition to DR integrated resonant strengths, we also inferred DR rate coefficients using Equation \eqref{eq:DRrates} at different plasma electron temperatures. 
{Such rate coefficients are important for collisional-radiative models of single-temperature or multi-temperature plasmas.}
Table~\ref{table:3} lists our experimentally inferred and theoretical DR rate coefficients from $S_\mathrm{EXP}$ together with those available in the OPEN-ADAS and AtomDB \citep{Zatsarinny2004,Foster2012} databases.
}

\begin{figure}
\centering
\includegraphics[clip=true,width=\columnwidth]{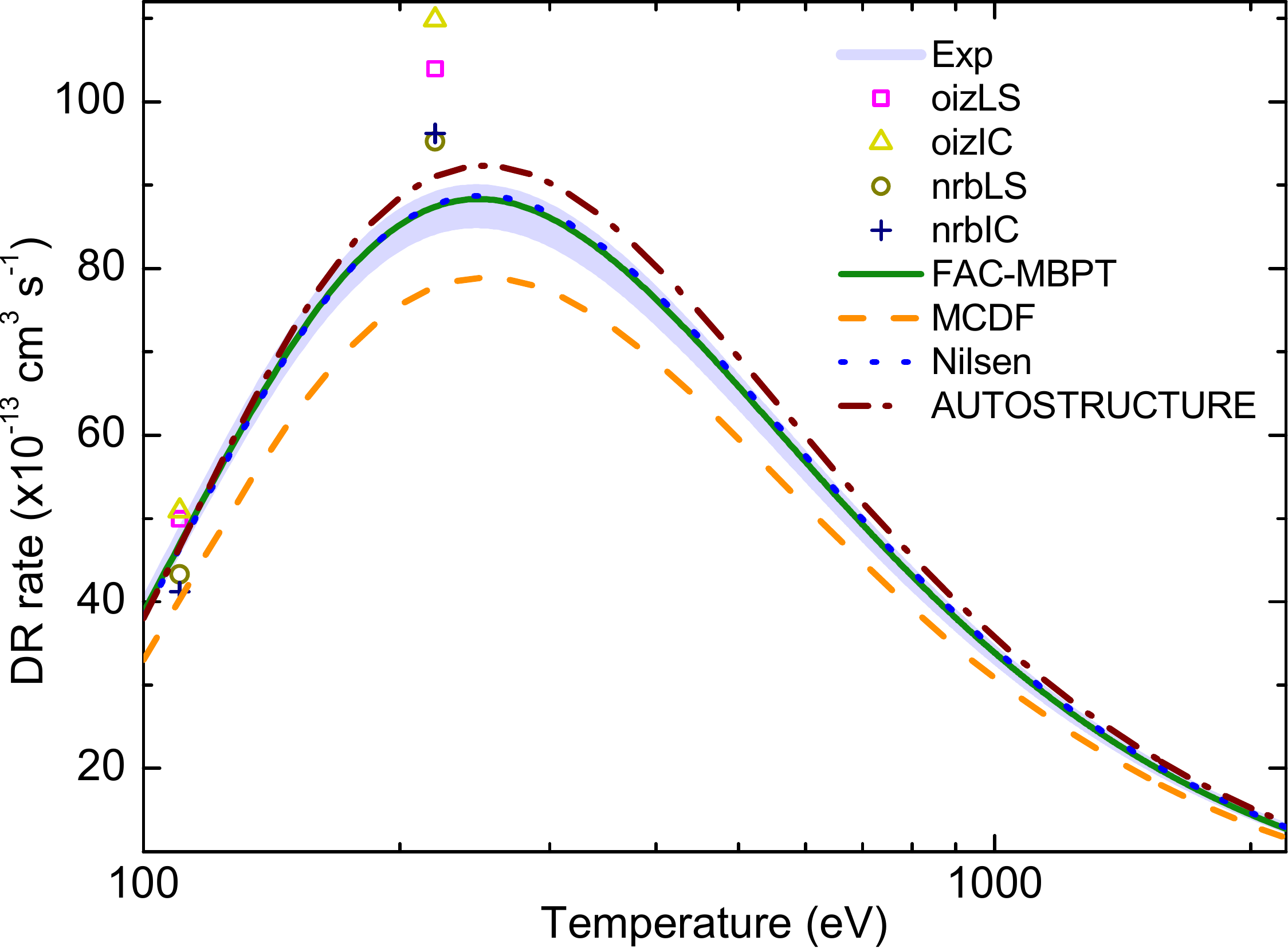} 
\caption{Experimental and theoretical DR rate coefficients  obtained with FAC-MBPT and MCDF and by~\citet{Nilsen1989} and \citet{Zatsarinny2004} (AUTOSTRUCTURE), as well as tabulated values for two temperatures from \texttt{adf09} files of OPEN-ADAS, namely oizLS, oizIC, nrbLS, and nrbIC.}
\label{fig:7}
\end{figure}

{
To compare with OPEN-ADAS, we identified the DR rates by the respective final-state configuration of the Na-like Fe ion. 
The combined total rates for the 3$s$, 3$p$, and 3$d$ final configurations are presented in Table ~\ref{table:3}. %
Since the DR rates of OPEN-ADAS include all DR LM$n$ channels, we restricted our comparison to the temperature range $T_i\leq 250$\,eV, where the LMM channel dominates ($\geq$ 95\% of the total decay into a given final-state configuration). 
Two tabulated DR rates in this temperature range ($T_{\mbox{e}}=110.3$~eV and 220.3~eV) were retrieved through \texttt{adf09} files, namely \texttt{oiz00\#ne\_fe16ls23} (oizLS) and \texttt{oiz00\#ne\_fe16ic23} (oizIC), provided by the author O.~Zatsarinny in LS coupling and IC, respectively, as well as \texttt{nrb00\#ne\_fe16ls23} (nrbLS) and \texttt{nrb00\#ne\_fe16ic23} (nrbIC) given by the author N. Badnell in respective atomic couplings. 
According to \citet{Foster2012}, the \ion{Fe}{17} DR rates in the AtomDB database were taken from \citet{Zatsarinny2004}, calculated using the AUTOSTRUCTURE code, and summed over all final states.

Since some minor final-state configurations are shared by the observed experimental spectral features, only the total experimental rates are presented in Table \ref{table:3}. 
Figure \ref{fig:7} depicts the comparison between our experiment and all available DR rates over a large temperature range.
Our experimental rate coefficients agree well with the theoretical rates within $\sim$0.2--0.8\% for those calculated with FAC-MBPT and witin $\sim$7--14\% for those calculated with MCDF methods. 
The DR rates taken from~\citet{Nilsen1989} are in close agreement with the FAC-MBPT calculations.
The OPEN-ADAS data provided by Badnell in both LS coupling and IC depart by $\sim-9$ to 12\% from our experimental results, while the data provided by Zatsarinny show differences of $\sim$6--25\%.
The total DR rates obtained with AUTOSTRUCTURE \citep{Zatsarinny2004} (also used in the AtomDB database) display differences from our results  of $<$5\% for the entire electron temperature range.
}

%
\section{Summary and Conclusions}
\label{sec:sum}

The DR LMM resonances for \ion{Fe}{17} ions have been measured using two different EBITs and compared with results obtained at the TSR \citep{Schmidt2009}. 
We simulated the time-dependent charge-state distribution to rule out systematic effects due to the presence of spurious charge states and depletion of charge states due to DR, which may affect our resonant strength measurements. 
{
The LMM DR integrated strengths were extracted from all three experiments, and compared with calculations performed using FAC, FAC-MBPT, and MCDF codes. 
Among them, FAC-MBPT calculations show the best agreement with the experiments. 
We also derived from our experimental data LMM DR rate coefficients for several temperatures. 
Our experimental rates show good agreement with rates obtained from~\citet{Nilsen1989} and FAC-MBPT, while rates available in the OPEN-ADAS and AtomDB databases, which are frequently used for the analysis of astrophysical spectra, show departures from our experimentally inferred values.
These differences highlight how crucial laboratory measurements are for testing different spectral models. 
}
This is important not only from the perspective of upcoming X-ray satellite missions XRISM \citep{xrism2018} and Athena \citep{Barret2016}, which will soon provide high-resolution spectra using X-ray microcalorimeters \citep{durkin2019}, but also for interpreting available high-resolution spectra from the operating Chandra and XMM-Newton observatories \citep{Gu2020} needed for reliable plasma diagnostics \citep{Beiersdorfer2014, Beiersdorfer2018}.

%
\vspace{4mm}
\section*{acknowledgments}
F.G. and P.A. acknowledge support from Funda\c{c}\~{a}o para a Ci\^{e}ncia e Tec\-no\-lo\-gia, Portugal, under grant No.~UID/FIS/04559/2020 (LIBPhys) and under contracts UI/BD/151000/2021 and SFRH/BPD/92329/2013. C.S. acknowledges support from an appointment to the NASA Postdoctoral Program at the NASA Goddard Space Flight Center, administered by the Universities Space Research Association under a contract with NASA, by Deutsche For\-schungs\-ge\-mein\-schaft project No.~266229290, and by Max-Planck-Gesellschaft. We also thank Dr. Stefan Schippers for providing the raw data of \ion{Fe}{17} DR rates measured at the TSR in Heidelberg, Germany.

\section*{Appendix \\ \vspace{2mm} DR Calculations for \ion{Fe}{17}}\label{sec:DR_cal}

Table~\ref{table:ener_Pos} contains the resonant energies and strengths for the main spectral features that are benchmarked in this work. The emitted wavelengths of the decay channels are listed in Table~\ref{table:WPos} with additional data provided by \citet{Nilsen1989} and \citet{Beiersdorfer2014}. 

{The complete set of resonant energies, strengths, and emitted wavelengths calculated within this work (FAC, FAC-MBPT, and MCDF methods) is available online as machine-readable tables. These data can be used to make Figure \ref{fig:2} as well as to produce the synthetic spectra shown in Figure \ref{fig:6}. }

\setlength{\tabcolsep}{4pt}
\renewcommand{\arraystretch}{1.2}
\begin{table*}
\caption{Theoretical Values of the Peak Resonant Energies $E$ (eV) and Strengths $S$ ($10^{-20}$cm$^2$eV) Obtained in This Work with FAC, FAC-MBPT, and MCDF.}
\centering
\begin{tabular*}{\textwidth}{c @{\extracolsep{\fill}} clllcccccc}
\hline
\hline
Label	&	Intermediate State	&		&	Final State	&	${E}_{\text{FAC}}$	&	${E}_{\text{MBPT}}$	&	${E}_{\text{MCDF}}$	&	${S}_{\text{FAC}}$	&	${S}_{\text{MBPT}}$	&	${S}_{\text{MCDF}}$	\\
\hline
	&	$[((2p^{2}_{1/2} 2p^{3}_{3/2})_{3/2} 3s_{1/2})_{2} 3d_{5/2}]_{1/2}$ 	&	${}^{4} D_{1/2}$	&	$3s_{1/2} {}^{2} S_{1/2}$	&	313.06	&	310.23	&	313.08	&	6.04	&	6.42	&	2.74	\\
\hline																			
$s_{1}$ 	&	$[((2p_{1/2} 2p^{4}_{3/2})_{1/2} 3s_{1/2})_{1} 3d_{5/2}]_{3/2}$ 	&	${}^{2} D_{3/2}$	&	$3s_{1/2} {}^{2} S_{1/2}$	&	325.70	&	322.21	&	321.98	&	31.91	&	29.54	&	7.34	\\
\hline																			
$s_{2}$	&	$[((2p_{1/2} 2p^{4}_{3/2})_{1/2} 3s_{1/2})_{1} 3d_{3/2}]_{3/2}$ 	&	${}^{2} P_{3/2}$	&	$3s_{1/2} {}^{2} S_{1/2}$	&	334.62	&	330.72	&	328.60	&	25.89	&	23.22	&	20.61	\\
\hline																			
	&	$[((2p^{2}_{1/2} 2p^{3}_{3/2})_{3/2} 3p_{1/2})_{2} 3d_{5/2}]_{3/2}$ 	&	 ${}^{4} P_{3/2}$	&	$3p_{1/2} {}^{2} P_{1/2}$	&	341.50	&	338.94	&	339.95	&	1.04	&	1.11	&	0.24	\\
	&	$[((2p^{2}_{1/2} 2p^{3}_{3/2})_{3/2} 3p_{3/2})_{0} 3d_{5/2}]_{5/2}$ 	&	${}^{2} D_{5/2}$	&	$3p_{3/2} {}^{2} P_{3/2}$	&	354.47	&	351.41	&	352.30	&	6.16	&	7.99	&	4.59	\\
	&	$[((2p_{1/2} 2p^{4}_{3/2})_{1/2} 3p_{1/2})_{1} 3d_{5/2}]_{5/2}$ 	&	${}^{4} F_{5/2}$	&	$3p_{3/2} {}^{2} P_{3/2}$	&	355.8	&	353.04	&	354.05	&	5.14	&	4.95	&	0.04	\\
\hline																			
$p_{1}$	&	$[((2p_{1/2} 2p^{4}_{3/2})_{1/2} 3p_{1/2})_{1} 3d_{3/2}]_{3/2}$ 	&	${}^{4} F_{3/2}$	&	$3p_{1/2} {}^{2} P_{1/2}$	&	353.65	&	355.05 	&	352.28	&	1.06	&	9.70	&	0.16	\\
	&	$[((2p_{1/2} 2p^{4}_{3/2})_{1/2} 3p_{1/2})_{1} 3d_{3/2}]_{3/2}$ 	&	${}^{4} F_{3/2}$	&	$3p_{3/2} {}^{2} P_{3/2}$	&	353.65	&	355.05	&	352.28	&	0.008	&	0.22	&	0.32	\\
	&	$[((2p_{1/2} 2p^{4}_{3/2})_{1/2} 3p_{3/2})_{2} 3d_{5/2}]_{5/2}$ 	&	${}^{2} D_{5/2}$	&	$3p_{3/2} {}^{2} P_{3/2}$	&	358.36	&	355.09	&	356.22	&	0.03	&	0.61	&	0.48	\\
	&	$[((2p_{1/2} 2p^{4}_{3/2})_{1/2} 3p_{1/2})_{1} 3d_{3/2}]_{3/2}$ 	&	${}^{2} D_{3/2}$	&	$3p_{1/2} {}^{2} P_{1/2}$	&	361.25	&	357.38	&	358.83	&	48.50	&	47.02	&	38.59	\\
	&	$[((2p^{2}_{1/2} 2p^{3}_{3/2})_{3/2} 3p_{3/2})_{2} 3d_{5/2}]_{5/2}$ 	&	${}^{2} D_{5/2}$	&	$3p_{3/2} {}^{2} P_{3/2}$	&	361.67	&	358.04	&	359.77	&	35.26	&	34.74	&	30.35	\\
	&	$[((2p_{1/2} 2p^{4}_{3/2})_{1/2} 3p_{1/2})_{1} 3d_{3/2}]_{1/2}$	&	${}^{2} S_{1/2}$	&	$3p_{1/2} {}^{2} P_{1/2}$	&	361.71	&	358.42	&	367.70	&	12.90	&	12.36	&	12.68	\\
\hline																			
$p_{2}$	&	$[(2s_{1/2} 2p^{2}_{1/2} 2p^{4}_{3/2})_{1/2} 3s^{2}_{1/2}]_{1/2}$ 	&	$\cdots$	&	$3p_{3/2} {}^{2} P_{3/2}$	&	364.85 	&	361.80 	&	$\cdots$	&	5.56	&	4.45	&	$\cdots$	\\
	&	$[((2p_{1/2} 2p^{4}_{3/2})_{1/2} 3p_{3/2})_{2} 3d_{3/2}]_{1/2}$  	&	${}^{2} P_{1/2}$	&	$3p_{3/2} {}^{2} P_{3/2}$	&	367.42	&	363.36	&	364.06	&	10.77	&	11.87	&	19.61	\\
	&	$[((2p_{1/2} 2p^{4}_{3/2})_{1/2} 3p_{3/2})_{1} 3d_{3/2}]_{5/2}$ 	&	${}^{2} D_{5/2}$	&	$3p_{3/2} {}^{2} P_{3/2}$	&	368.99	&	364.74	&	371.33	&	8.69	&	4.43	&	29.69	\\
	&	$[((2p_{1/2} 2p^{4}_{3/2})_{1/2} 3p_{3/2})_{1} 3d_{5/2}]_{3/2}$ 	&	${}^{2} D_{3/2}$	&	$3p_{3/2} {}^{2} P_{3/2}$	&	369.50	&	365.35	&	366.12	&	12.17	&	11.39	&	16.64	\\
																			
\hline																			
$d_{1}$	&	$[((2p^{2}_{1/2} 2p^{3}_{3/2})_{3/2} 3d_{3/2})_{2} 3d_{5/2}]_{5/2}$ 	&	${}^{4} G_{5/2}$	&	$3d_{3/2} {}^{2} D_{3/2}$	&	396.39	&	393.3	&	396.73	&	9.42	&	11.46	&	11.37	\\
	&	$[((2p^{2}_{1/2} 2p^{3}_{3/2})_{3/2} 3d^{2}_{5/2}]_{7/2}$ 	&	${}^{4} D_{7/2}$	&	$3d_{5/2} {}^{2} D_{5/2}$	&	397.93	&	394.65	&	396.22	&	11.89	&	13.96	&	12.55	\\
\hline																			
$d_{2}$	&	$[((2p_{1/2} 2p^{4}_{3/2})_{1/2} 3d^{2}_{5/2}]_{7/2}$  	&	${}^{2} G_{7/2}$	&	$3d_{5/2} {}^{2} D_{5/2}$	&	404.14	&	400.95	&	402.46	&	13.45	&	12.08	&	11.92	\\
	&	$[((2p_{1/2} 2p^{4}_{3/2})_{1/2} 3d_{3/2})_{2} 3d_{5/2}]_{5/2}$ 	&	${}^{2} F_{5/2}$	&	$3d_{3/2} {}^{2} D_{3/2}$	&	404.45	&	401.05	&	402.21	&	10.19	&	24.19	&	1.33	\\
	&	$[((2p_{1/2} 2p^{4}_{3/2})_{1/2} 3d_{3/2})_{2} 3d_{5/2}]_{5/2}$ 	&	${}^{2} F_{5/2}$	&	$3d_{5/2} {}^{2} D_{5/2}$	&	404.45	&	401.05	&	402.21	&	7.00	&	6.27	&	1.78	\\
	&	$[((2p_{1/2} 2p^{4}_{3/2})_{1/2} 3d^{2}_{3/2}]_{5/2}$ 	&	${}^{2} F_{5/2}$	&	$3d_{3/2} {}^{2} D_{3/2}$	&	405.22	&	401.98	&	403.14	&	41.48	&	23.62	&	35.91	\\
\hline																			
$d_{3}$	&	$[((2p_{1/2} 2p^{4}_{3/2})_{1/2} 3d^{2}_{5/2}]_{1/2}$  	&	${}^{2} P_{1/2}$	&	$3d_{3/2} {}^{2} D_{3/2}$	&	411.16	&	407.57	&	408.61	&	6.25	&	7.53	&	9.59	\\
	&	$[((2p_{1/2} 2p^{4}_{3/2})_{1/2} 3d_{3/2})_{1} 3d_{5/2}]_{7/2}$	&	${}^{2} F_{7/2}$	&	$3d_{5/2} {}^{2} D_{5/2}$	&	411.90	&	407.96	&	410.07	&	60.63	&	58.13	&	52.73	\\
	&	$[((2p_{1/2} 2p^{4}_{3/2})_{1/2} 3d_{3/2})_{1} 3d_{5/2}]_{5/2}$ 	&	${}^{2} D_{5/2}$	&	$3d_{5/2} {}^{2} D_{5/2}$	&	412.85	&	408.75	&	410.29	&	14.74	&	13.79	&	16.84	\\
\hline																			
	&	$[((2p_{1/2} 2p^{4}_{3/2})_{1/2} 3d_{3/2})_{1} 3d_{5/2}]_{3/2}$ 	&	${}^{2} P_{3/2}$	&	$3d_{3/2} {}^{2} D_{3/2}$	&	415.20	&	411.12	&	412.39	&	12.26	&	12.66	&	17.84	\\
	&	$[((2p_{1/2} 2p^{4}_{3/2})_{1/2} 3d_{3/2})_{1} 3d_{5/2}]_{3/2}$ 	&	${}^{2} P_{3/2}$	&	$3d_{5/2} {}^{2} D_{5/2}$	&	415.20	&	411.12	&	412.39	&	15.55	&	16.92	&	18.24	\\
	&	$[((2p_{1/2} 2p^{4}_{3/2})_{1/2} 3d^{2}_{3/2}]_{1/2}$  	&	${}^{2} P_{1/2}$	&	$3d_{3/2} {}^{2} D_{3/2}$	&	420.27	&	416.73	&	415.67	&	6.01	&	5.39	&	7.49	\\

\hline
\end{tabular*}\\
\vspace{1ex}
{\raggedright \textbf{Note.} The resonant and final states are given in j-j and LSJ notations. \par}

\label{table:ener_Pos}
\end{table*}

\setlength{\tabcolsep}{4pt}
\renewcommand{\arraystretch}{1.2}
\begin{table*}
\caption{Theoretical Values of Emitted Wavelengths (\AA) Obtained in This Work with FAC, FAC-MBPT, and MCDF.}
\centering
\begin{tabular*}{\textwidth}{c @{\extracolsep{\fill}} clllccccc}
\hline
\hline
Label	&	Intermediate State	&		&	Final State	&	$\lambda_{\text{FAC}}$	&	$\lambda_{\text{FAC-MBPT}}$	&	$\lambda_{\text{MCDF}}$	&	\citet{Nilsen1989}	&  \citet{Beiersdorfer2014}	\\
\hline
	&	$[((2p^{2}_{1/2} 2p^{3}_{3/2})_{3/2} 3s_{1/2})_{2} 3d_{5/2}]_{1/2}$ 	&	${}^{4} D_{1/2}$	&	$3s_{1/2} {}^{2} S_{1/2}$	&	15.47	&	15.52	&	15.46	&	15.52	&	15.49	\\
\hline																	
s$_{1}$ 	&	$[((2p_{1/2} 2p^{4}_{3/2})_{1/2} 3s_{1/2})_{1} 3d_{5/2}]_{3/2}$ 	&	${}^{2} D_{3/2}$	&	$3s_{1/2} {}^{2} S_{1/2}$	&	15.23	&	15.29	&	15.29	&	15.28	&	15.27	\\
\hline																	
s$_{2}$	&	$[((2p_{1/2} 2p^{4}_{3/2})_{1/2} 3s_{1/2})_{1} 3d_{3/2}]_{3/2}$ 	&	${}^{2} P_{3/2}$	&	$3s_{1/2} {}^{2} S_{1/2}$	&	15.06	&	15.13	&	15.17	&	15.12	&	15.11	\\
\hline																	
	&	$[((2p^{2}_{1/2} 2p^{3}_{3/2})_{3/2} 3p_{1/2})_{2} 3d_{5/2}]_{3/2}$ 	&	 ${}^{4} P_{3/2}$	&	$3p_{1/2} {}^{2} P_{1/2}$	&	15.59	&	15.63	&	15.61	&	15.58	&	$\cdots$	\\
	&	$[((2p^{2}_{1/2} 2p^{3}_{3/2})_{3/2} 3p_{3/2})_{0} 3d_{5/2}]_{5/2}$ 	&	${}^{2} D_{5/2}$	&	$3p_{3/2} {}^{2} P_{3/2}$	&	15.38	&	15.44	&	15.42	&	15.63	&	$\cdots$	\\
	&	$[((2p_{1/2} 2p^{4}_{3/2})_{1/2} 3p_{1/2})_{1} 3d_{5/2}]_{5/2}$ 	&	${}^{4} F_{5/2}$	&	$3p_{3/2} {}^{2} P_{3/2}$	&	15.36	&	15.41	&	15.39	&	15.41	&	15.41	\\
\hline																	
p$_{1}$ &	$[((2p_{1/2} 2p^{4}_{3/2})_{1/2} 3p_{1/2})_{1} 3d_{3/2}]_{3/2}$ 	&	${}^{4} F_{3/2}$	&	$3p_{1/2} {}^{2} P_{1/2}$	&	15.35	&	15.32	&	15.37	&	15.40	&	15.30	\\
	&	$[((2p_{1/2} 2p^{4}_{3/2})_{1/2} 3p_{1/2})_{1} 3d_{3/2}]_{3/2}$ 	&	${}^{4} F_{3/2}$	&	$3p_{3/2} {}^{2} P_{3/2}$	&	15.40	&	15.38	&	15.42	&	15.45	&	$\cdots$	\\
	&	$[((2p_{1/2} 2p^{4}_{3/2})_{1/2} 3p_{3/2})_{2} 3d_{5/2}]_{5/2}$ 	&	${}^{2} D_{5/2}$	&	$3p_{3/2} {}^{2} P_{3/2}$	&	15.26	&	15.37	&	15.35	&	15.37	&	$\cdots$	\\
	&	$[((2p_{1/2} 2p^{4}_{3/2})_{1/2} 3p_{1/2})_{1} 3d_{3/2}]_{3/2}$ 	&	${}^{2} D_{3/2}$	&	$3p_{1/2} {}^{2} P_{1/2}$	&	15.21	&	15.27	&	15.25	&	15.26	&	15.26	\\
	&	$[((2p^{2}_{1/2} 2p^{3}_{3/2})_{3/2} 3p_{3/2})_{2} 3d_{5/2}]_{5/2}$ 	&	${}^{2} D_{5/2}$	&	$3p_{3/2} {}^{2} P_{3/2}$	&	15.25	&	15.31	&	15.28	&	15.55	&	15.29	\\
	&	$[((2p_{1/2} 2p^{4}_{3/2})_{1/2} 3p_{1/2})_{1} 3d_{3/2}]_{1/2}$	&	${}^{2} S_{1/2}$	&	$3p_{1/2} {}^{2} P_{1/2}$	&	15.20	&	15.26	&	15.08	&	15.25	&	15.07	\\
\hline																	
p$_{2}$	&	$[(2s_{1/2} 2p^{2}_{1/2} 2p^{4}_{3/2})_{1/2} 3s^{2}_{1/2}]_{1/2}$ 	&	$\cdots$	&	$3p_{3/2} {}^{2} P_{3/2}$	&	15.19	&	15.24	&	$\cdots$	&	15.23	&	15.24	\\
	&	$[((2p_{1/2} 2p^{4}_{3/2})_{1/2} 3p_{3/2})_{2} 3d_{3/2}]_{1/2}$  	&	${}^{2} P_{1/2}$	&	$3p_{3/2} {}^{2} P_{3/2}$	&	15.14	&	15.21	&	15.20	&	15.19	&	15.19	\\
	&	$[((2p_{1/2} 2p^{4}_{3/2})_{1/2} 3p_{3/2})_{1} 3d_{3/2}]_{5/2}$ 	&	${}^{2} D_{5/2}$	&	$3p_{3/2} {}^{2} P_{3/2}$	&	15.11	&	15.18	&	15.07	&	15.17	&	$\cdots$	\\
	&	$[((2p_{1/2} 2p^{4}_{3/2})_{1/2} 3p_{3/2})_{1} 3d_{5/2}]_{3/2}$ 	&	${}^{2} D_{3/2}$	&	$3p_{3/2} {}^{2} P_{3/2}$	&	15.10	&	15.17	&	15.16	&	15.20	&	15.16	\\
																	
\hline																	
d$_{1}$	&	$[((2p^{2}_{1/2} 2p^{3}_{3/2})_{3/2} 3d_{3/2})_{2} 3d_{5/2}]_{5/2}$ 	&	${}^{4} G_{5/2}$	&	$3d_{3/2} {}^{2} D_{3/2}$	&	15.48	&	15.53	&	15.47	&	15.54	&	15.50	\\
	&	$[((2p^{2}_{1/2} 2p^{3}_{3/2})_{3/2} 3d^{2}_{5/2}]_{7/2}$ 	&	${}^{4} D_{7/2}$	&	$3d_{5/2} {}^{2} D_{5/2}$	&	15.45	&	15.51	&	15.49	&	15.52	&	15.49	\\
\hline																	
d$_{2}$	&	$[((2p_{1/2} 2p^{4}_{3/2})_{1/2} 3d^{2}_{5/2}]_{7/2}$  	&	${}^{2} G_{7/2}$	&	$3d_{5/2} {}^{2} D_{5/2}$	&	15.34	&	15.39	&	15.37	&	15.40	&	15.37	\\
	&	$[((2p_{1/2} 2p^{4}_{3/2})_{1/2} 3d_{3/2})_{2} 3d_{5/2}]_{5/2}$ 	&	${}^{2} F_{5/2}$	&	$3d_{3/2} {}^{2} D_{3/2}$	&	15.32	&	15.38	&	15.36	&	15.38	&	$\cdots$	\\
	&	$[((2p_{1/2} 2p^{4}_{3/2})_{1/2} 3d_{3/2})_{2} 3d_{5/2}]_{5/2}$ 	&	${}^{2} F_{5/2}$	&	$3d_{5/2} {}^{2} D_{5/2}$	&	15.33	&	15.39	&	15.37	&	15.39	&	$\cdots$	\\
	&	$[((2p_{1/2} 2p^{4}_{3/2})_{1/2} 3d^{2}_{3/2}]_{5/2}$ 	&	${}^{2} F_{5/2}$	&	$3d_{3/2} {}^{2} D_{3/2}$	&	15.31	&	15.36	&	15.35	&	15.37	&	15.35	\\
\hline																	
d$_{3}$	&	$[((2p_{1/2} 2p^{4}_{3/2})_{1/2} 3d^{2}_{5/2}]_{1/2}$  	&	${}^{2} P_{1/2}$	&	$3d_{3/2} {}^{2} D_{3/2}$	&	15.20	&	15.26	&	15.24	&	15.26	&	15.24	\\
	&	$[((2p_{1/2} 2p^{4}_{3/2})_{1/2} 3d_{3/2})_{1} 3d_{5/2}]_{7/2}$	&	${}^{2} F_{7/2}$	&	$3d_{5/2} {}^{2} D_{5/2}$	&	15.19	&	15.26	&	15.22	&	15.25	&	15.23	\\
	&	$[((2p_{1/2} 2p^{4}_{3/2})_{1/2} 3d_{3/2})_{1} 3d_{5/2}]_{5/2}$ 	&	${}^{2} D_{5/2}$	&	$3d_{5/2} {}^{2} D_{5/2}$	&	15.17	&	15.24	&	15.22	&	15.23	&	15.22	\\
\hline																	
	&	$[((2p_{1/2} 2p^{4}_{3/2})_{1/2} 3d_{3/2})_{1} 3d_{5/2}]_{3/2}$ 	&	${}^{2} P_{3/2}$	&	$3d_{3/2} {}^{2} D_{3/2}$	&	15.12	&	15.19	&	15.17	&	15.35	&	15.18	\\
	&	$[((2p_{1/2} 2p^{4}_{3/2})_{1/2} 3d_{3/2})_{1} 3d_{5/2}]_{3/2}$ 	&	${}^{2} P_{3/2}$	&	$3d_{5/2} {}^{2} D_{5/2}$	&	15.13	&	15.20	&	15.18	&	15.36	&	15.18	\\
	&	$[((2p_{1/2} 2p^{4}_{3/2})_{1/2} 3d^{2}_{3/2}]_{1/2}$  	&	${}^{2} P_{1/2}$	&	$3d_{3/2} {}^{2} D_{3/2}$	&	15.03	&	15.09	&	15.11	&	15.09	&	15.08	\\

\hline
\end{tabular*}\\
\vspace{1ex}
{\raggedright \textbf{Notes.} The resonant and final states are given in $j$-$j$ and LSJ notations. The respective data provided by \citet{Nilsen1989} and by \citet{Beiersdorfer2014} are also listed for comparison. \par}
\label{table:WPos}
\end{table*}



\end{document}